\documentclass[a4paper,11pt]{article}
\pdfoutput=1 

\usepackage{jheppub} 
\usepackage[T1]{fontenc} 
\usepackage{appendix}

\usepackage{amssymb,amsfonts}
\usepackage{amsmath}
\usepackage{mathtools}
\usepackage{amsthm}
\usepackage{empheq}
\usepackage[colorinlistoftodos]{todonotes}
\usepackage{array}

\newcommand{\be}{\begin{equation}}
\newcommand{\ee}{\end{equation}}
\newcommand{\bea}{\begin{eqnarray}}
\newcommand{\eea}{\end{eqnarray}}

\def\Dslash{\,\,{\raise.15ex\hbox{/}\mkern-13mu D}}
\def\Dbarslash{\,\,{\raise.15ex\hbox{/}\mkern-12mu {\bar D}}}
\def\delslash{\,\,{\raise.15ex\hbox{/}\mkern-10mu \partial}}
\def\delbarslash{\,\,{\raise.15ex\hbox{/}\mkern-9mu {\bar\partial}}}
\def\pslash{\,\,{\raise.15ex\hbox{/}\mkern-11mu p}}
\def\qslash{\,\,{\raise.15ex\hbox{/}\mkern-9mu q}}
\def\kslash{\,\,{\raise.15ex\hbox{/}\mkern-11mu k}}
\def\eslash{\,\,{\raise.15ex\hbox{/}\mkern-9mu \epsilon}}

\newcommand{\slsh}[1]{\,\,{\raise.15ex\hbox{/}\mkern-12mu {#1}}}

\setcitestyle{square}

\title{\boldmath High-frequency Gravitational Waves from Superstring Phases in the Early Universe}

\author[a]{Joseph P. Conlon,}
\author[b]{Edmund J. Copeland,}
\author[a]{Edward Hardy,}
\author[a]{Noelia S\a'anchez Gonz\a'alez}

\affiliation[a]{Rudolf Peierls Centre for Theoretical Physics,\\Beecroft Building,\\ Parks Rd, Oxford, OX1 3PU,\\UK}
\affiliation[b]{School of Physics and Astronomy, \\
University of Nottingham, \\
Nottingham, \\
NG7 2RD, UK
}

\emailAdd{joseph.conlon@physics.ox.ac.uk}
\emailAdd{ed.copeland@nottingham.ac.uk}
\emailAdd{edward.hardy@physics.ox.ac.uk}
\emailAdd{noelia.sanchezgonzalez@physics.ox.ac.uk}

\preprint{November 2025}

\abstract{When moduli roll in the early universe, all physical scales -- including string tensions -- simultaneously evolve. The dynamics of cosmic string loops with time-varying tension can produce cosmic string loop trackers in which most of the energy density of the universe lies in the form of string loops. This solution can exist as an attractor until the rolling modulus reaches its minimum, when the loops ultimately decay through gravitational wave emission. We explore the spectrum of gravitational waves produced by such string loop trackers. The resulting spectrum is high-frequency and peaks in the GHz regime today. The amplitude of the signal is diluted by any subsequent matter-dominated epochs, and thus the potential observability of the signal crucially depends on the duration of the moduli-dominated epoch that follows once the moduli settle down and oscillate about their minimum.}

\begin{document} 
\maketitle
\flushbottom
\section{Introduction and Review}\label{Sec1}
The very early universe is arguably the best location to look for 
observational effects of string theory (for general reviews of string cosmology, see \cite{Cicoli:2023opf, Brandenberger:2023ver}). Probes such as gravitational waves offer new windows on the universe and there are few current constraints on the state of the universe between the epochs of inflation (probed by the CMB) and nucleosynthesis (probed by primordial element abundances).
\\
\\
This allows for equations of state that differ from simple radiation domination. In \cite{SanchezGonzalez:2025uco}, we recently identified a novel attractor solution in which three-quarters of the energy density of the universe is in the form of loops of cosmic superstrings (see \cite{Brunelli:2025eif} for a recent paper generalising 
this to include effective strings from wrapped NS5 branes). This result arose from considering the dynamics of strings with a time-varying tension. Cosmic superstrings, first proposed in \cite{Witten:1985fp}, can have time-varying tensions as their tension is set by the vevs of the compactification moduli, which in the early universe may be displaced from the current vacuum. 
\\
\\
As the volume modulus rolls towards the asymptotic boundary of moduli space,\footnote{Such an evolution is strongly motivated in the context of the LARGE Volume Scenario \cite{Balasubramanian:2005zx,Conlon:2005ki}: as the final volume, $\mathcal{V}$, of the extra-dimensions is at exponentially large values, $\mathcal{V} \gg 1$ in string units, the earlier post-inflationary cosmological evolution of the volume modulus must lead to the minimum (for details see \cite{Apers:2024ffe}). During this evolution, the tension of F-strings decreases and loops of F-strings grow with time as seen in Eq.~\eqref{length growth}.} the string tension, $\mu$, reduces (as $\mu \sim m_s^2 \sim M_P^2/\mathcal{V}$), where $m_s$ is the fundamental string scale and $M_P$ is the four-dimensional Planck mass. As the tension reduces, the radius of sub-horizon string loops, $\ell(t)$, increases with time \cite{Conlon:2024uob} (also see \cite{Revello:2024gwa, Brunelli:2025ems, Ghoshal:2025tlk}, and for older work on strings with time-varying tensions, see \cite{Yamaguchi:2005gp, Ichikawa:2006rw, Cheng:2008ma, Sadeghi:2009wx, Wang:2012naa, Emond:2021vts}),
\begin{align}\label{length growth}
    \ell(t) = \ell_i \sqrt{\frac{\mu_i}{\mu(t)}},
\end{align}
where $\ell_i$ and $\mu_i$ are the loops' initial length and tension respectively. This is in contrast with the constant tension case, when sub-horizon loops of cosmic strings slowly evaporate due to emission of gravitational radiation.
Furthermore, as the volume modulus is rolling, the combination of increasing length and decreasing tension makes string loops stable against evaporation from gravitational wave emission \cite{Conlon:2024uob}. 
\\
\\
Given the canonically normalised volume modulus $\Phi$, defined as\footnote{In string theory, this arises from a K\"ahler potential $K = - 3 \ln \left( T + \bar{T} \right)$ and $\mathcal{V} \sim {\textrm{Re}}(T)^{2/3}$, where $T$ is the complexified K\"ahler modulus.}
\begin{equation}
    \Phi = \sqrt{\frac{2}{3}}M_P \ln \mathcal{V},
\end{equation}
the energy density of a population of loops of fundamental superstrings can be written as a function of the scale factor $a$ and the  modulus $\Phi$, 
\begin{equation}
    \rho_{\rm loops}(a,\Phi) = n \cdot \mu \cdot \ell =\rho_{\rm loops, i} \cdot \left(\frac{a_i}{a}\right)^3 \cdot e^{-\sqrt{3/8} (\Phi - \Phi_i)/M_P},
\end{equation}
where $n$ is the number density of loops at any given scale factor $a$, $\rho_{\rm loops, i}$ is the energy density of the initially formed loops with scale factor $a_i$ and modulus $\Phi_i$ is the initial modulus vev.
\\
\\
One of the main results of \cite{SanchezGonzalez:2025uco} was a new attractor solution for the volume modulus with an LVS potential $V\sim \mathcal{V}^{-3}$, in which most of the energy of the universe lies in the form of loops of fundamental superstrings (therefore called a string loop tracker). The solution remains an attractor so long as the modulus is rolling. On it, the physical radius of loops increases but the comoving radius does not, so a population of initially isolated loops remain isolated.\footnote{As proposed in \cite{Conlon:2024uob}, another possibility due to a time-dependent tension is for the population of loops to percolate and form a network that includes infinitely long strings during an early kination epoch. We do not focus on that scenario here.}  
Once the modulus starts oscillating around its minimum, the loops begin to decay through emission of gravitational waves. In our present work, we compute the spectrum of these gravitational waves. 
We do not specify the origin of the initial loop population, which could arise through e.g. nucleation during inflation or reheating at the end of D-brane inflation. Moduli oscillations end with moduli-driven reheating and the beginning of the Hot Big Bang era.
\\
\\
The paper is structured as follows: in Section~\ref{Sec3} we compute the gravitational wave signal from this new population of loops, with the important aspect that they decay while moduli are oscillating about their minimum. In Section~\ref{sec4} we consider specific string motivated scenarios for the modulus dominated epoch and show how the signal encodes information about the physics of those scenarios. We conclude in Section~\ref{sec5}.

\section{Gravitational Waves from Cosmic String Loops}\label{Sec3}

\subsection{Summary of the Physics}\label{summarysec}
In this subsection we summarise the physics of the gravitational wave emission and the various epochs that are present.
\begin{figure}[ht!]
    \centering
    \includegraphics[scale=0.47]{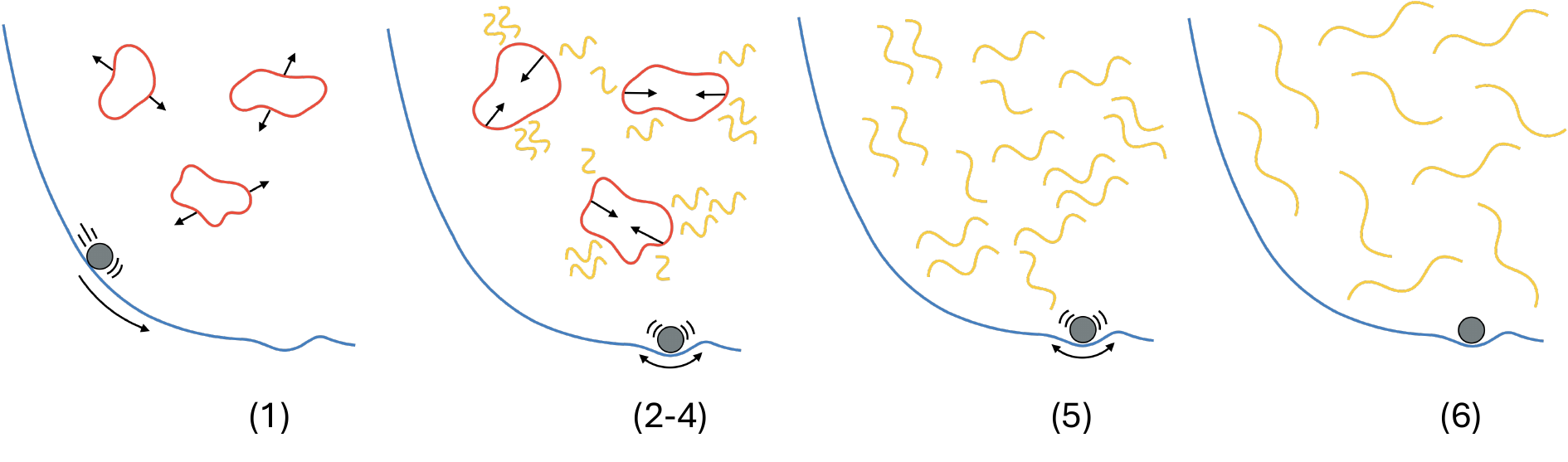}
    \caption{Different phases of the evolution of the system we consider, which  ultimately leads to a stochastic background of gravitational waves (yellow sinusoidal waves) that were emitted by cosmic string loops (red) as the modulus (grey ball) evolves towards the minimum of its potential (blue). The numbers refer to the epochs discussed in Section~\ref{summarysec}.} 
    \label{fig:placeholder}
\end{figure}
\begin{enumerate}

\item The initial conditions are those of the string loop tracker. In this phase, 75\% of the energy density of the universe lies in the form of cosmic string loops, and the remaining 25\% of the energy density is in the form of the kinetic energy of the rolling scalar field. In this epoch the kinetic energy of the rolling scalar is dominant over its potential. 

\item The second epoch occurs as the modulus approaches the minimum of the potential. In the earlier, rolling phase, the potential is a decaying exponential. However, as the modulus approaches the minimum the shape of the potential turns into an (approximate) quadratic, with a minimum at vanishing vacuum energy.\footnote{The vanishing vacuum energy has to arise from tuning, given the lack of good explanations for the value of the cosmological constant.}

\item As the modulus begins to oscillate around the minimum, the universe enters a matter dominated epoch. The oscillating modulus behaves as dust and redshifts as matter. The loops of cosmic string also initially behave as matter. So long as they are separated and do not intersect, their number density redshifts as $a^{-3}$. In this epoch, cosmic strings and oscillating moduli co-exist with comparable energy densities, both redshifting as matter.

\item Meanwhile, the cosmic strings are slowly emitting gravitational radiation. The characteristic frequency of this emission is set by the string length, $f^{\rm em} = 2/\ell$. As soon as such radiation is generated, it begins to redshift relative to the background matter distribution and its frequency decreases.

\item Once the cosmic strings decay completely into gravitational radiation, the universe enters a moduli-dominated epoch. During this era the amplitude and frequency of the gravitational waves both redshift as dark radiation (meaning it will never interact with Standard Model degrees of freedom, but it will still contribute to the overall energy density of the universe).

\item The duration of the moduli dominated epoch is set by the moduli lifetime. Once the moduli eventually decay, they reheat the universe and the ordinary Hot Big Bang commences. From this point onwards, the gravitational radiation continues to behave as dark radiation. 
\end{enumerate}

\subsection{Gravitational wave power spectrum}
The gravitational wave power spectrum measured today is given by the density parameter
\begin{equation}
    \Omega^0_{\rm GW}(f^{\rm obs}) = \frac{1}{\rho_{\rm c,0}} \frac{d\rho_{\rm GW}(t_0)}{d\log f^{\rm obs}},
\end{equation}
where $\rho_{\rm GW}(t_0)$ is the energy density in gravitational waves today, $\rho_{\rm c,0}$ is today's critical density and $f^{\rm obs}$ denotes the observed frequency. $\Omega^0_{\rm GW}$ corresponds to the energy density in gravitational waves per logarithmic frequency interval, normalised by the critical energy density today. Given the energy density emitted per unit of time $\dot{\rho}_{\rm GW}^{\rm em}$, it is related to the energy density observed today $\rho_{\rm GW}(t_0)$ through an integral over the duration of the emission (between times that we call $t_i$ to $t_f$),
\begin{equation}\label{PSintegral}
     \Omega^0_{\rm GW}(f^{\rm obs}) = \int_{t_i}^{t_f} dt \; \frac{f^{\rm obs}}{\rho_{c,0}} \frac{d\dot{\rho}^{\rm em}_{\rm GW}(f^{\rm em},t)}{df^{\rm em}}  \left(\frac{a(t)}{a_0}\right)^3,
\end{equation}
where the frequency of the gravitational waves at the time of emission is related to the frequency observed today by $f^{\rm em} = f^{\rm obs} \cdot  a_0/a(t)$. 
Here we have used the fact that the energy density in gravitational radiation redshifts proportional to $a^{-4}$.  As mentioned earlier, we are interested in the spectrum of gravitational waves emitted by a population of loops after moduli start oscillating about their minimum. We therefore take 
$t_{\rm i}$, the start of gravitational wave emission, to be the corresponding time.\footnote{Gravitational waves are also emitted by the string loops prior to the moduli oscillating, while they are still rolling. As will be clear, the present-day energy density from such times is strongly diluted and observationally less important than that from the era we consider.}
\\
\\
A standard cosmic string loop with constant tension $\mu$ decays due to gravitational radiation, emitted with a power set by its tension, $dE_{\rm GW}/dt= \Gamma G\mu^2$ where $\Gamma \simeq 50$ according to Nambu-Goto simulations \cite{Vilenkin:2000jqa}.  There may be additional model-dependent decay channels, but here we focus on the case in which gravitational emission is the main decay route.\footnote{The loops can also lose energy by emitting the modulus $\Phi$ via a coupling of the schematic form $S_{\rm int} \sim -\int d^2\xi\, \mu(\Phi_0) \frac{\tilde{\Phi}}{M_{\rm Pl}} \sqrt{-\gamma}$, where $\tilde{\Phi}=\Phi-\Phi_0$ denotes small fluctuations around  the homogeneous background $\Phi_0$ ($\xi$ are the world-sheet coordinates and $\gamma$ is the induced metric). For a typical oscillating loop, the power emitted in $\Phi$ is parametrically $dE_\Phi/dt\sim G \mu^2$, as for gravitational waves, so this will affect our results by at most an order-one factor.}  As a result, the loop's length evolves with time as follows:
\begin{equation}\label{length evolution}
    \ell(t)=\ell_i - \Gamma G\mu (t-t_i),
\end{equation}
where $\ell_i$ is the initial length at the initial time $t_i$. Consequently, one can define the lifetime of the loop as $\tau=\ell_i/(\Gamma G\mu)$. Furthermore, the possible frequencies of emission are the different harmonics $k$ set by the size of the loop, 
\begin{equation}\label{femitted}
    f^{\rm em}_k = \frac{2k}{\ell}\; , \; k \in \mathbb{N},
\end{equation}
with $k=1$ corresponding to the fundamental mode. The power efficiency associated to each mode $\Gamma^{(k)}$ is related to the total power efficiency defined above through $\Gamma^{(k)} \propto k^{-4/3}  \Gamma $ \cite{Damour:2000wa,Damour:2001bk}. For an accessible review of the computation of the gravitational power spectrum from cosmic strings we direct the reader to \cite{Gouttenoire:2022gwi}.
\\
\\
The radiation emitted by a population of isolated cosmic string loops consists of the superposition of the different emissions from the individual loops. We will describe the population by the number density as a function of the loop length and time, $n_{\rm loops}(\ell,t)$. The loop population will be diluted due to cosmic expansion, such that $n_{\rm loops} \propto a^{-3}$. Here, we do not consider the effects of loop production that would be taking place if there was a network of strings present, instead focusing on a population of isolated loops.
\\
\\
Therefore, the energy density emitted in gravitational waves per unit of time and per unit of frequency can be written as
\begin{equation}
 \frac{d\dot{\rho}_{\rm GW}^{\rm em}(t)}{df^{\rm em}} = \sum_k  \frac{dE^{(k)}_{\rm GW}}{dt} \cdot \frac{d n_{\rm loops}(\ell(k,f^{\rm em}),t)}{df^{\rm em}} ,
\end{equation}
where $dE_{\rm GW}^{(k)}/dt$ denotes the power emitted by the harmonic $k$ of a single loop. Given a certain frequency of emission and a certain harmonic $k$, the length of the corresponding loop, $\ell(k,f^{\rm em})$, is fixed  by Eq.~\eqref{femitted}. 
As a first approximation, we take all the loops to have the same length $\ell$, such that 
\begin{equation}\label{rhoemitted}
    \frac{d\dot{\rho}_{\rm GW}^{\rm em}(t)}{df^{\rm em}} = \sum_k \Gamma^{(k)}G\mu^2 \cdot n_{\rm loops}(t^{\rm em}) \cdot \delta\left(f^{\rm em} - \frac{2k}{\ell(t^{\rm em})}\right),
\end{equation}
where we have substituted $dE^{(k)}_{\rm GW}/dt = \Gamma^{(k)} G\mu^2$. The delta function above imposes that given a harmonic $k$ and a frequency of emission $f^{\rm em}$, there is a corresponding length, and thus a corresponding time of emission. One should keep in mind the frequency emitted is a function of a given frequency observed through the appropriate redshift factor, $f^{\rm em}= f^{\rm obs} \cdot a_0/a(t^{\rm em})$.
\\
\\
We could consider different profiles for the distribution in loop length, which would affect the profile of the gravitational wave signal, widening its frequency spectrum as different sizes correspond to different frequencies of emission. The appropriate distribution of lengths would be given by the formation mechanism of the population  -- for examples of recent work in this direction see \cite{Blanco_Pillado_2014},
\cite{Auclair:2025jtd}, and in terms of the effect the distributions have on the stochastic gravitational wave signal LIGO could detect, see \cite{LIGOScientific:2025kry}. However, as we remain agnostic about this, when presenting results in the following sections, we will consider the simplest possibility with all the loops having the same size. This also amounts to focusing on the fraction of the population whose length dominates the emission.
\\
\\
Introducing Eq.~\eqref{rhoemitted} in Eq.~\eqref{PSintegral} and using the properties of the delta function, one arrives at the following expression,
\begin{equation}
       \Omega^0_{\rm GW} (f^{\rm obs})= \int_{t_i}^{t_f} dt \; \frac{1}{\rho_{c,0}}  \cdot \sum_k \left[ \Gamma^{(k)}G\mu^2 \cdot n_{\rm loops}(t^{\rm em}) \; \frac{\delta(t-t^{\rm em}(f^{\rm obs}, k))}{\left|\dot{\ell}/\ell - \dot{a}/a\right|_{t=t^{\rm em}}}\right]  \; \left(\frac{a(t)}{a_0}\right)^4,
\end{equation}
where the time of emission $t^{\rm em}$ is a function of the frequency observed $f^{\rm obs}$ for a given $k$ via
\begin{equation}\label{freqobs}
   f^{\rm obs} = \frac{2k}{\ell(t^{\rm em})} \frac{a(t^{\rm em})}{a_0}.
\end{equation}
Evaluating the integrand at the time of emission  $t^{\rm em}$ defined implicitly in Eq.~\eqref{freqobs} and using $\dot{\ell} = - \Gamma G\mu$, the gravitational power spectrum simplifies to
\begin{equation}\label{PSformula}
   \Omega^0_{\rm GW} (f^{\rm obs})= \sum_k \left[{\Omega^0_{\rm GW,i}}^{(k)} \cdot \frac{H_i}{\left(H(t^{\rm em})+ \frac{\Gamma G\mu}{\ell(t^{\rm em})}\right)} \cdot \left(\frac{a(t^{\rm em})}{a_i}\right)\right],
\end{equation}
where $H=\dot{a}/a$ is the Hubble parameter and the subscript $i$ refers to the start of moduli oscillations. The first factor ${\Omega^0_{\rm GW,i}}^{(k)}$ appearing in Eq.~\eqref{PSformula} is given by
\begin{equation}\label{amplitudeFirstHubble}
  {\Omega^0_{\rm GW,i}}^{(k)} = \frac{\Gamma^{(k)} G\mu }{\varepsilon}\cdot \Omega^i_{\rm loops} \cdot\left(\frac{H_i}{H_0}\right)^2 \cdot \left(\frac{a_i}{a_0}\right)^4,
\end{equation}
where $\varepsilon=\ell_i H_i$ is the initial size of the loops in horizon units and
\begin{equation}\label{initialloopdensity}
\Omega^i_{\rm loops}= \frac{\mu \;\ell_i\;n_i}{\rho_{c,i}}
\end{equation}
denotes the initial fraction of the total energy density stored in loops, where $\rho_{c,i}$ is the critical energy density of the universe once moduli start oscillating about their minimum. This factor $\Omega_{\rm GW,i}^0$ corresponds to the contribution to the power spectrum coming from the emission in the first Hubble time, i.e. $t^{\rm em}\sim H_i^{-1}$. 
\\
\\
The prefactor in Eq.~\eqref{PSformula}, ${\Omega^0_{\rm GW,i}}^{(k)}$ defined in Eq.~\eqref{amplitudeFirstHubble}, has a straightforward physical interpretation. First, one realises the quantity $ \Gamma G\mu/\varepsilon=(dE_{\rm GW}/dt\cdot H_i^{-1})/(\mu \ell_i)$ is the fraction of the initial energy of a loop emitted in gravitational waves in the first Hubble time. Moreover,  $\Omega_{\rm loops}^i$ is the fraction of the energy of the universe initially stored in the loops such that $\Omega_{\rm loops}^i \cdot 3M_P^2 H_i^2$ is the total energy density of the loops at the start of moduli oscillations. As a result, $\Gamma G\mu/\varepsilon \cdot \Omega_{\rm loops}^i \cdot 3M_P^2 H_i^2$ is the energy density in gravitational waves that is produced in the first Hubble time. 
\\
\\
On the other hand, the factor  $(a_i/a_0)^4$ accounts for the redshift of the gravitational radiation due to the cosmic expansion from the moment of emission until it reaches us today. Finally, dividing by $3M_P^2H_0^2$ gives us the ratio of that gravitational emission in the first Hubble time with respect to the energy density of the universe today, which is what we have denoted as $\Omega^0_{\rm GW,i}$.

\subsection{Amplitude and frequency of the peak today}
The gravitational power spectrum from the population of loops will have a peak. Its corresponding amplitude and frequency can be computed analytically given the simplifications made in the previous section where only one length for the loop population is considered. We will focus on the first harmonic $k=1$ to get an estimate of the orders of magnitude, rather than an exact expression. Looking at the turning points of the expression obtained for the power spectrum Eq.~\eqref{PSformula} as a function of frequency $f^{\rm obs}$, one finds the peak corresponds to a time of emission
\begin{equation}\label{tempeak}
    t^{\rm em}_{\rm peak} =\frac{3\sqrt{11}-7}{5} \cdot \left(\frac{\ell_i}{\Gamma G\mu} + t_i\right) \simeq 0.6 \; \tau,
\end{equation}
where $\tau=\ell_i/\Gamma G\mu \gg t_i$ is the lifetime of the loops, which is much longer than a Hubble time provided the loops, although sub-horizon, are sufficiently large with $\varepsilon=\ell_iH_i \gg  G\mu$. The numerical factor in Eq.~\eqref{tempeak} is obtained by assuming the emission happens while moduli are oscillating, such that $a \sim t^{2/3}$. 
The physical interpretation of Eq.~\eqref{tempeak} is clear: the peak amplitude corresponds to emission just before the loops decay away.
\\
\\
Substituting the time of peak emission into Eq.~\eqref{freqobs}, one can determine the frequency observed corresponding to the peak,
\begin{align}\label{fpeakanalytic}
    f_{\rm peak}^{\rm obs} \simeq \;\ell^{-1}_i \cdot \frac{a_{\rm rh}}{a_0} \cdot \frac{a_{\tau}}{a_{\rm rh}} \simeq \;\ell^{-1}_i \cdot \frac{a_{\rm rh}}{a_0} \cdot \left(\frac{\tau}{t_{\rm rh}}\right)^{2/3}.
\end{align}
where $t_{\rm rh}$ denotes the end of moduli domination, when the modulus decays and reheating takes place. 
\\
\\ 
The amplitude at the peak corresponds to the energy in the loops getting released mostly at the end of their lifetime and then redshifting until the present day. It can be likewise obtained by substituting the value from Eq.~\eqref{tempeak} into Eq.~\eqref{PSformula},
\begin{equation}\label{analytic amplitude v1}
\Omega^0_{\rm GW, peak} \simeq \Omega^0_{\rm GW,i}\cdot\frac{H(t_{\rm i})}{H(\tau)} \cdot \left(\frac{a(\tau)}{a (t_{\rm i})}\right),
\end{equation}
where $t_i$ denotes the start of the modulus oscillations, $\tau$ is the cosmic string loops lifetime and $\Omega^0_{\rm GW,i}$ is the contribution from the emission corresponding to the first Hubble time given in Eq.~\eqref{amplitudeFirstHubble}. Using the definition for $\Omega^0_{\rm GW,i}$ in Eq.~\eqref{amplitudeFirstHubble} and that their lifetime is given by $\tau = \ell_i/\Gamma G\mu$, the expression above can be rewritten as
\begin{equation}\label{analytic amplitude v2}
    \Omega^0_{\rm GW,peak} \simeq \Omega_{\rm loops}^i \left(\frac{H_{\rm i}}{H_0}\right)^2 \left(\frac{a(t_{\rm i})}{a_0}\right)^4  \left(\frac{a(\tau)}{a(t_{\rm i})}\right) .
\end{equation}
Finally, if we introduce that the fraction of energy density in radiation today is the total energy of the universe at reheating diluted by the cosmic expansion, $\Omega_{\rm rad}^0 = (H(t_{\rm rh})/H_0)^2 \cdot  (a(t_{rh})/a_0)^4$, and use the fact that during moduli domination $a\sim t^{2/3}$, we arrive at
\begin{equation} \label{analytic amplitude v3}
    \Omega^0_{\rm GW,peak} \simeq \Omega_{\rm loops}^i\cdot\Omega_{\rm rad}^0 \cdot  \left(\frac{H(t_{\rm rh})}{H(\tau)}\right)^{2/3}
\end{equation}
which, as expected, shows that the amplitude at the peak is basically the energy in loops at the start of moduli oscillations redshifted from the moment of decay to the present day.

\subsubsection{Parameter dependence}
We now connect the properties of the peak of the gravitational wave spectrum to the parameters defining our scenario: the initial size of the loops with respect to the horizon $ \varepsilon=\ell_iH_i$, their tension $G\mu$, along with the mass $m_\phi$ and decay rate $\Gamma_\phi$ of the modulus. Once the modulus starts oscillating, the theory settles down at its final vacuum and parameters such as the string tension become fixed and cease to evolve.
\\
\\
Moduli start oscillating once the Hubble scale becomes comparable to the mass of the modulus, i.e. $H_i \sim m_\phi$ . Plugging this into Eq.~\eqref{tempeak} we see that the time when the gravitational waves at the peak of the spectrum are emitted goes as $t^{\rm em}_{\rm peak} \simeq \tau = \varepsilon/(\Gamma G\mu \cdot m_\phi)$, while  from Eq.~\eqref{femitted} the corresponding emitted frequency is given by $f_{\rm peak}^{\rm em} \simeq \ell_i^{-1} =m_\phi /\varepsilon$. 
\\
\\
On the other hand, reheating and initiation of the radiation-dominated Hot Big Bang occurs when the Hubble parameter becomes comparable to the decay rate of the modulus, i.e. $H_{\rm rh} \sim \Gamma_\phi$.  We will parametrise the modulus decay rate as $\Gamma_\phi \equiv (\gamma/48\pi )\cdot m_\phi^3 /M_P^2$, where for the canonical scenario of moduli decay via gravitationally suppressed couplings $\gamma \sim \mathcal{O}(1)$ \cite{Cicoli:2012aq}.
\\
\\
Recalling that the temperature falls like $1/a$ due to the cosmic expansion, we can determine the scale factor at reheating $ a_{\rm rh}=a_0 \cdot (T_0/T_{\rm rh})$ where $T_0$ denotes the temperature today and the temperature of reheating is given by $T_{\rm rh} \sim \sqrt{M_P H_{\rm rh}}$. As a result, using Eq.~\eqref{fpeakanalytic} to relate the frequency emitted to the one observed, we obtain
\begin{align}\label{fpeakparam}
    &f_{\rm peak}^{\rm obs} \simeq T_0 \cdot  \left(\frac{1}{\varepsilon } \sqrt{\frac{\gamma}{48\pi} }\right)^{1/3} \left(\frac{1}{ \Gamma 
G\mu}\right)^{2/3}\left(\frac{ m_\phi}{M_P}\right)^{5/6}.
\end{align}
From Eq.~\eqref{fpeakparam} we see that the observed \emph{frequency} of the peak will be lower if:
\begin{itemize}
    \item The modulus, $m_\phi$, is lighter, which implies a longer modulus domination epoch and therefore, more redshift of the gravitational radiation during this early matter domination era.
    \item The modulus decay rate constant $\gamma$  is smaller, meaning a slower decay and thus, a longer moduli domination era and additional redshift.
    \item The loops are longer initially, meaning larger $\varepsilon$, as that implies a lower fundamental frequency of emission.
    \item  The loops have a larger tension $G\mu$, as this implies a shorter lifetime. The earlier the time of emission, the more the emitted gravitational waves are redshifted  before reaching us today.
\end{itemize}
We now turn our attention to the amplitude of the signal at the peak. Recall the lifetime of the loops $\tau$ is set by their initial size $\ell_i = \varepsilon / m_\phi $ and their power of emission $P_{GW} = \Gamma G\mu^2$ such that the final time of emission is $H_f^{-1} \sim \tau = \ell_i/(\Gamma G\mu)$. Consequently, from Eq.~\eqref{analytic amplitude v3} the amplitude depends on the parameters of the theory as follows,
\begin{align}\label{amplitudepeakparam}
   \Omega^0_{\rm GW,peak} \simeq \Omega_{\rm rad}^0 \;\Omega_{\rm loops}^i \; \left(\frac{\varepsilon}{\Gamma G\mu}\frac{\gamma}{48\pi} \right)^{2/3} \left(\frac{m_\phi}{M_P}\right)^{4/3}.
\end{align}
The \emph{amplitude} of the signal will therefore be larger if:
\begin{itemize}
    \item The modulus, $m_\phi$, is heavier, since that would mean a shorter moduli domination epoch and less dilution of the energy in gravitational radiation during this early matter dominated era.
    \item  The modulus decay rate constant $\gamma$  is larger which, again, means a faster decay and thus, a shorter moduli domination era with less dilution of the signal.
    \item The loops are longer initially, meaning larger $\varepsilon$, as that means a longer lifetime and less extra dilution of the energy stored in them since they redshift as matter just like the background.
    \item The loops have a smaller tension $G\mu$, which again implies a longer lifetime. Recall the loops redshift as matter and so, the longer they survive into the era of oscillating moduli before releasing their energy into gravitational radiation which does redshift with respect to the background during this matter dominated era, the less dilution the amplitude of the signal observed today will have suffered. 
\end{itemize}
\subsubsection{Numerical values}
Having obtained expressions for the frequency and amplitude of the peak, Eqs.~\eqref{fpeakparam} and~\eqref{amplitudepeakparam}, we can estimate their typical values. We have that some parameters are fixed, such as the coefficient in the emitted power $\Gamma \simeq 50$, as estimated by numerical simulations \cite{Vilenkin:2000jqa}. In addition, we have $M_P\simeq 2.4 \cdot 10^{18} \; \rm GeV$, $\Omega_{\rm rad}^0 \simeq 9.0\cdot 10^{-5}$ and $T_0 \simeq 2.4 \cdot 10^{-4} \; \rm eV \simeq 5.7 \cdot 10^{10}\; {\rm Hz}$. Certain parameters can also be estimated as order one, such as $\Omega_{\rm loops}^i \sim 1$ (recall in our scenario the population of loops is a significant fraction of the total energy density of the universe at the start of moduli oscillations, as we are coming off a string loop tracker). 
\\
\\
In the theories that we consider the loop population decays before Big Bang Nucleosynthesis (BBN), and we assume there are no infinite strings. As a result, many observational constraints on the tension of cosmic strings do not apply (these include, e.g., bounds from analysis of anisotropies in the CMB to be $G\mu < 10^{-7}$ \cite{Charnock:2016nzm} and more model-dependent bounds from searches for stochastic gravitational wave backgrounds with the Pulsar Timing Array, $G\mu \lesssim 10^{-10}$ \cite{LIGOScientific:2021nrg, Avgoustidis:2025svu, Ellis:2023tsl}, and from the LIGO-Virgo-KAGRA gravitational wave detectors, $G\mu \lesssim 10^{-15}$ \cite{LIGOScientific:2025kry}).
\\
\\
The longest-lived modulus must, however, decay before BBN takes place, as the equation of state of the universe at that point is constrained to correspond to radiation domination. This is also known as the (Strong) Cosmological Moduli Problem \cite{Coughlan:1983ci,Banks:1993en,deCarlos:1993wie}, and it implies $m_{\phi} \gtrsim 30 \; \rm TeV$ \cite{Blumenhagen:2009gk} for moduli decaying conventionally via gravitationally suppressed channels.
\\
\\
Putting numbers into Eqs. \eqref{amplitudepeakparam} and \eqref{fpeakparam}, the amplitude of the peak associated to this emission can be written as,
\begin{align}\label{amplitudenum}
   \Omega^0_{\rm GW,peak}  \simeq 10^{-19} \cdot \left( \varepsilon \cdot  \gamma\right)^{2/3} \cdot  \left( \frac{10^{-10}}{ G\mu }\right)^{2/3}  \cdot \left(\frac{m_\phi}{10 \; \rm TeV}\right)^{4/3},
\end{align}
while the frequency associated to the peak of this emission is given by
\begin{equation}\label{freqnum}
    f_{\rm peak}^{\rm obs} \simeq 10^{4} \;\text{Hz} \cdot \left(\frac{\sqrt{\gamma}}{\varepsilon}\right)^{1/3} \cdot \left(\frac{10^{-10}}{ G\mu}\right)^{2/3}\cdot\left(\frac{ m_\phi}{10\; \rm TeV}\right)^{5/6}.
\end{equation}
For our approximations and analysis to be consistent, there is an upper bound on the parameter $\varepsilon$.  
Naively, one would think that for sub-horizon loops the bound should be $\varepsilon = \ell_iH_i <1$ (corresponding to horizon size loops). However, the nature of the string loop tracker found in \cite{SanchezGonzalez:2025uco} implies that $\varepsilon$ will actually be smaller. This arises from the way the tracker has an $\mathcal{O}(1)$ fractional energy density in string loops which -- although they grow in absolute size -- shrink relative to the horizon while the modulus evolves. In the tracker, the population of loops forms an order one fraction of the total energy density at the start of moduli oscillations and the loops are initially isolated from each other (which sets an upper bound on their size given a certain number density). 
\\
\\
In particular, assuming the loops are an $\mathcal{O}(1)$ fraction of the total energy density, the Friedmann equation gives $N_H \sim 1/(G\mu \cdot \varepsilon)$, where $N_H$ is the number of loops per Hubble volume and recall $\varepsilon = \ell_i H_i$. In order to avoid the impact of interactions, the number density has to be lower than one per the unit volume occupied by a loop, i.e. $N_H < 1/\varepsilon^3$. As a result, it follows that the tracker has $\varepsilon < \sqrt{G\mu}$. Introducing this in the expressions for the amplitude \eqref{amplitudenum} and the frequency \eqref{freqnum}, one obtains
\begin{align} 
\Omega^0_{\rm GW,peak}  &\lesssim 10^{-23} \cdot  \gamma^{2/3} \cdot  \left( \frac{10^{-10}}{ G\mu }\right)^{1/3}  \cdot \left(\frac{m_\phi}{10 \; \rm TeV}\right)^{4/3}\label{peakbounds1}
\\
f_{\rm peak}^{\rm obs} &\gtrsim 10^5 \; {\rm Hz}\cdot \gamma^{1/6} \cdot \left(\frac{10^{-10}}{G\mu } \cdot \frac{m_\phi}{10 \; \rm TeV}\right)^{5/6} .\label{peakbounds2}
\end{align}
These bounds come from restricting to isolated sub-horizon loops. Recall a decreasing tension makes sub-horizon cosmic string loops stable against gravitational emission if they are above a critical size. In \cite{SanchezGonzalez:2025uco}, the string loop tracker solution was obtained neglecting interactions and assuming loops were initially above that critical size. However, if we allow for interactions, loop inter-commutations might produce loops small enough to evaporate quickly via gravitational emission. In that case, a more careful analysis is needed to determine whether the population of sub-horizon loops can become an order one fraction of the total energy density.
\\
\\
Although the amplitude in Eq.~\eqref{peakbounds1} appears tiny, there is hope. The smallness of the amplitude arises from the length of the moduli-dominated matter epoch. As seen in Eq.~\eqref{analytic amplitude v3}, the least amount of dilution is achieved if the loops decay at the end of the modulus oscillations or after, that is $\tau\gtrsim  t_{\rm rh}$, in which case, the amplitude becomes $\Omega^0_{\rm GW, peak} \gtrsim  \Omega^0_{\rm rad}$. One should keep in mind dark radiation bounds \cite{Planck:2018vyg}, which imply any additional relativistic degrees of freedom (such as present gravitational radiation) should be subdominant to the ones coming from the Standard Model, such that $\Omega_{\rm GW, peak}^0 \lesssim 10^{-1} \Omega_{\rm rad}^0$.  
\\
\\
Therefore, if the loops are a significant fraction of the energy density of the universe at the start of moduli oscillations, as it would be if they were to enter this epoch from a string loop tracker as the one presented in \cite{SanchezGonzalez:2025uco}, their gravitational radiation can saturate the dark radiation bound if they survive until the end of this matter dominated era.  In principle, to achieve this, the loops should either have a greater lifetime, due to a lower tension or bigger initial size, or the moduli domination epoch should be shorter, due to a faster decay rate or a heavier modulus mass. Success in shortening the modulus lifetime will achieve an increase in the gravitational wave amplitude. In Section~\ref{sec4}, we will present their gravitational power spectrum considering different possibilities for the decay rate of the modulus. 

\section{Specific Scenarios for the Modulus Epoch}\label{sec4}

As emphasised above, the amplitude for the signal depends crucially on the modulus decay rate. In this section, we study the gravitational wave signal in the context of different modulus decay scenarios; as discussed, the faster this decay the larger the amplitude of the present-day gravitational wave signal.
\\
\\
We first consider a canonical modulus decay rate, which leads to a long modulus domination era due to the modulus's gravitationally suppressed couplings. Next, we present the gravitational wave power spectrum in the context of a faster decay, made possible due to an enhanced coupling to the Higgs sector in the context of fine-tuned high-scale SUSY \cite{Cicoli:2022fzy}. In the last part of this section, we discuss the signal that would be observed if the modulus decay were able to occur through non-gravitational couplings (although note that the microphysics that would lead to this in the context of string theory compactifications is unknown). If the route was possible, this last scenario would constitute the best opportunity for signals with large amplitudes. 
\\
\\
The LARGE Volume Scenario (LVS) \cite{Balasubramanian:2005zx,Conlon:2005ki}
realises the idea of large extra dimensions in the context of type IIB flux compactifications where all moduli can be stabilised at exponentially large volume in string units, $\mathcal{V} \gg 1$. It allows us to naturally build hierarchies at the same time as guaranteeing control of the effective field theory expansion in $\alpha'$. In the context of a large volume compactification, the parameters that appear in our description are set by the vev of the volume modulus.  
\\
\\
Large volumes resulting in a lower string scale, allow for cosmic super-strings with low tensions $G\mu \ll 1$ (necessary for compatibility with observational bounds). In the case of F-strings their tension depends on the volume as  $\mu_F = m_s^2 \sim M_P^2/\mathcal{V}$. Low tensions can also be achieved through warping in the extra dimensions, although the specific dependence on the compactification details becomes more model-dependent. In addition to this, the mass of the volume modulus is also set by the volume of the compactification as $(m_\phi/M_P) \sim \mathcal{V}^{-3/2}$~\cite{Conlon:2005ki}. 

\subsection{The Large Volume Scenario with a canonical volume modulus}
As just mentioned, in LVS we find a hierarchy of masses among which there is a light volume modulus, $(m_\phi/M_P) \sim \mathcal{V}^{-3/2}$. One of the most general predictions of string cosmology is the necessity of an epoch of moduli domination as these light scalar fields oscillate about the minimum in which they eventually  stabilize (see \cite{Cicoli:2023opf, Brandenberger:2023ver} for recent reviews in string cosmology). 
\\
\\
The fact that moduli interactions are Planck-suppressed makes them long-lived, which in turn, since radiation redshifts faster than matter and their decay products can be expected to be relativistic, motivates the idea that eventually the longest-lived massive field comes to dominate the energy of the universe before reheating to the Standard Model degrees of freedom. 
\\
\\
In a vanilla LVS, the decay rate of the light volume modulus is set by Planck-suppressed interactions, such that
\begin{equation}\label{canonicaldecay}
    \Gamma_{\phi}^{\rm (Vanilla \; LVS)} =\frac{\gamma}{48\pi} \frac{m_\phi^3}{M_P^2},
\end{equation}
where $\gamma \sim \mathcal{O}(1)$. The branching ratio to its own axion is the same order as to the SM via the Higgses. This gives rise to a dark radiation overproduction problem \cite{Cicoli:2012aq}. In the next section, we consider the faster decay route presented in \cite{Cicoli:2022fzy} which allows for a solution to this problem.
\\
\\
This decay route Eq.~\eqref{canonicaldecay} in which $\gamma \sim \mathcal{O}(1)$, leads in Eqs.~\eqref{peakbounds1} and \eqref{peakbounds2} to 
\begin{align} 
\Omega^0_{\rm GW,peak}  &\lesssim 10^{-23} \cdot  \left( \frac{10^{10}}{\mathcal{V}}\right)^{5/3} \label{amplitude bound grav}
\\
f_{\rm peak}^{\rm obs} &\gtrsim 10^{5} \; {\rm Hz} \cdot \left( \frac{10^{10}}{\mathcal{V}}\right)^{5/12}, \label{freq bound grav}
\end{align}
where we have used the dependence of the tension and the mass of the modulus on the volume that correspond to LVS. As expected we see a very suppressed amplitude due to the late modulus decay and the dilution of radiation during this long moduli dominated era. 

\subsection{The Large Volume Scenario with a fast-decaying volume modulus}
\begin{figure}[t!]
    \centering
    \includegraphics[scale=0.7]{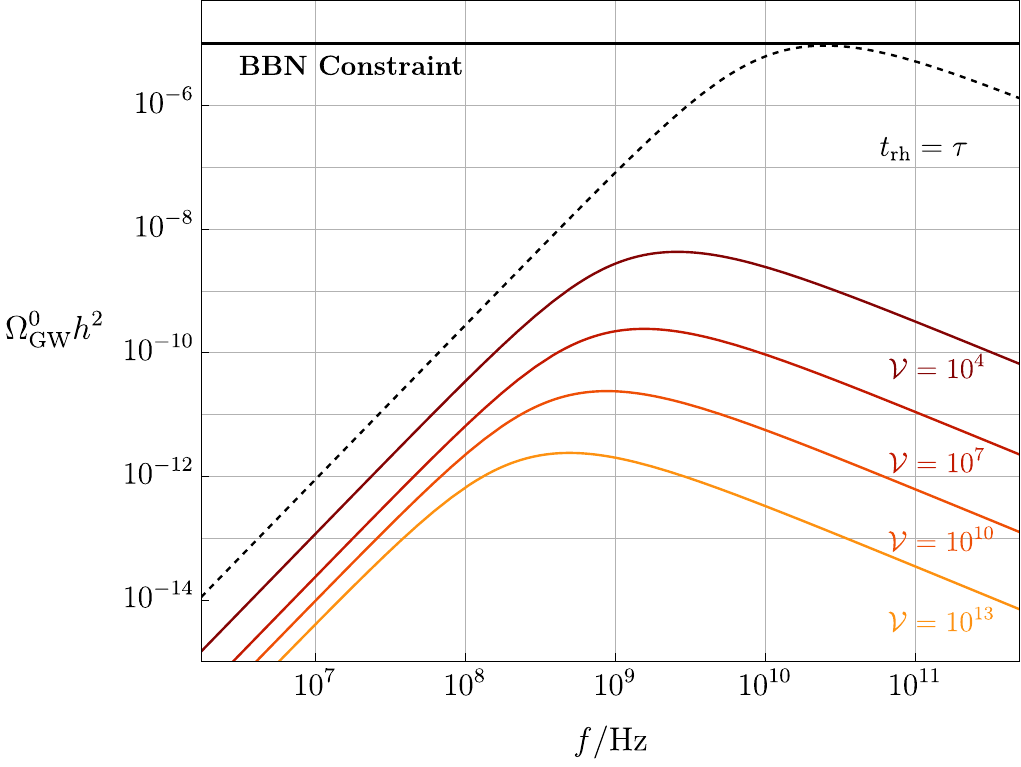} \qquad\qquad \qquad
    \caption{Gravitational power spectrum from a population of loops emitting while the volume modulus is oscillating about its minimum. The coloured lines correspond to emission in the context of LVS and a fast modulus decay rate $\Gamma_\phi \simeq (1/4\pi)^2(M_P/m_\phi)^{4/3} \cdot m_\phi^3/ M_P^2$. The dashed line represents the best case scenario, realised in the case of modulus decay via non-gravitational couplings, allowing for the time of reheating $t_{\rm rh}$ to occur at the end of the lifetime of the loops $\tau$. In the plot, this is shown for a volume in string units of $\mathcal{V}=10^{10}$ with $\Omega_{\rm loops}^{\rm i}$ taken such that the peak saturates the BBN constraint. The initial size of the loops is taken to be $\varepsilon=\sqrt{G\mu}$.}
    \label{fig:plotGWforLVS}
\end{figure}

We now consider a scenario presented in \cite{Cicoli:2022fzy} which allows for a faster decay of the volume modulus. This short volume lifetime (SVL-LVS) will allow for significantly larger amplitudes in the gravitational wave power spectrum in comparison with the vanilla LVS, as previously discussed in \cite{Ghoshal:2025tlk}. Additionally, this reheating scenario is well-motivated since it avoids the, previously mentioned, dark radiation constraints.
\\
\\
We start by summarising the scenario of \cite{Cicoli:2022fzy}. In the context of high-energy SUSY breaking, the observed electroweak scale, which in turn sets the Higgs mass, has to be fine tuned. Typically the electroweak scale gets a correction from SUSY-breaking contributions of order of the gravitino mass $m_{3/2}^2\sim M_P/\mathcal{V}$, so we can write schematically
\begin{equation}
    m_H^2 \sim m_{3/2}^2 \left[c_0 + c_{\rm loop} \log \left(\frac{m_{KK}}{m_{3/2}}\right)\right] ,
\end{equation}
where $c_0$ sets the bare mass and  $c_{\rm loop} \sim 1/4\pi$. The logarithmic enhancement appearing above is due to the running from the Kaluza Klein (KK) scale $m_{KK}$ down to the SUSY-breaking scale. Due to the required fine-tuned cancellation of the terms in brackets such that $m_H^2 \ll m_{3/2}^2$, when expanding to linear order in fluctuations of the canonically normalised volume modulus $\phi =\sqrt{3/2} \; M_P  \log{\mathcal{V}}$, the effect of expanding the logarithm $\log (m_{KK}/m_{3/2}) \sim \log (\sqrt{\mathcal{V}})$ dominates. As a result, what enters in the coupling of the Higgs $h$ to the light volume modulus $\delta \phi$ is the natural untuned mass, 
\begin{equation}
    \mathcal{L}_{\rm int} \sim \left(c_{\rm loop} m_{3/2}^2\right) h^2 \frac{\delta \phi}{M_P} .
\end{equation}
Then, parametrically we have $\Gamma_{\phi \rightarrow hh} \sim c_{\rm loop}^2 m_{3/2}^4/(m_{\phi} M_P^2)$, and Eq.~\eqref{canonicaldecay} is enhanced with respect to the branching ratio to its own axion by a factor $(m_{3/2}/m_\phi)^4 \sim \mathcal{V}^2$, such that we can write
\begin{equation}\label{SVLDecayRate}
    \Gamma_{\phi}^{\rm(SVL-LVS)} \simeq  (c_{\rm loop} \mathcal{V})^2 \frac{m_\phi^3}{M_P^2},
\end{equation}
This gives a shorter volume lifetime and can help with avoiding dark radiation constraints \cite{Cicoli:2022fzy}. In a particular compactification, one would have to check other moduli present which could outlive this one and produce dark radiation as well. 
\\
\\
Introducing this enhancement which, comparing the form of Eq.~\eqref{canonicaldecay} with Eq.~\eqref{SVLDecayRate}, corresponds to $\gamma/48\pi \sim (\mathcal{V}/4\pi)^2$, in Eq.~\eqref{amplitudepeakparam} results in
\begin{equation} \label{SVLAmplitude}
     \Omega^0_{\rm GW,peak}  \simeq \Omega_{\rm rad}^0 \Omega^i_{\rm loops} \;  \left(\frac{\varepsilon}{(4\pi)^2\Gamma}\right)^{2/3} \simeq 10^{-7}\; \varepsilon^{2/3},
\end{equation}
where we see it depends on the size of the loops in horizon units at the start of moduli oscillations $\varepsilon=\ell_i H_i$. Meanwhile, the frequency associated to the peak after substituting this fast decay rate in Eq.~\eqref{peakbounds2} is
\begin{equation}
  f_{\rm peak}^{\rm obs}\simeq 10^{7}\; \text{Hz} \cdot \varepsilon^{-1/3}\cdot \left(\frac{10^{10}}{\mathcal{V}}\right)^{1/4}.
 \end{equation}
Introducing the bound discussed above on the initial size of the loops in order that interactions between them are negligible,  $\varepsilon <\sqrt{G\mu} \sim 1/\sqrt{\mathcal{V}}$, the amplitude and frequency of the peak satisfy
\begin{align}
\Omega^0_{\rm GW,peak}  &\lesssim 10^{-10} \; \left( \frac{10^{10}}{\mathcal{V}}\right)^{1/3} \label{amplitude bound SVL}
\\
f_{\rm peak}^{\rm obs} &\gtrsim 10^8 \; {\rm Hz} \;\left( \frac{10^{10}}{\mathcal{V}}\right)^{1/12} ,\label{freq bound SVL}
\end{align}
which are saturated for the longest allowed isolated loops that constitute an order-one fraction of the total energy density at the start of moduli oscillations.  
\\
\\
The absence of a direct dependence on the volume in the expression for the amplitude found in Eq.~\eqref{SVLAmplitude} can be explained by recalling Eq.~\eqref{analytic amplitude v3}, where the amplitude is set by the ratio between the decay time of cosmic string loops $\tau$ and the time when reheating occurs $t_{\rm rh}$. In the case of a short volume lifetime in the Large Volume Scenario (SVL-LVS), this ratio only depends on the size of the loops at the start,
\begin{align}
\frac{\tau}{t_{\rm rh}} \sim \frac{\varepsilon}{\Gamma G\mu \cdot m_\phi} \; \left(\frac{M_P}{m_\phi}\right)^{4/3}\; \frac{m_\phi^3}{M_P^2} \sim \frac{\varepsilon}{\Gamma} ,
\end{align}
as $G\mu\sim \mathcal{V}^{-1}$ and $m_\phi/M_P \sim \mathcal{V}^{-3/2}$.
Since the loops are initially sub-horizon, such that $\varepsilon \ll 1$, we have extra dilution from their decay before reheating. 
\\
\\
In Figure~\ref{fig:plotGWforLVS}, we present the gravitational power spectrum for a population of isolated loops which corresponds to a significant fraction of the total energy at the start of moduli oscillations in the SVL-LVS case, with the volume modulus decay rate given in Eq.~\eqref{SVLDecayRate}. The dashed line labelled $t_{\rm rh}=\tau$ corresponds to the best case scenario achieved through a non-gravitationally suppressed decay, which we discuss in the next section. 
\subsection{Non-gravitational decays}
It is reasonable to also consider a scenario in which the modulus decays far more rapidly than considered above. The idea of a non-gravitational decay refers to a scenario where the modulus decays extremely rapidly, although at the moment the microphysics that would allow a string theory modulus to decay so rapidly is not known (given its origin as a gravitationally coupled degree of freedom). Nevertheless we include this as a potential possibility as it can lead to the largest observational signal.
\\
\\
Alternatively, one could forgo the connection between our work and string theory and instead consider a purely four-dimensional field theory model. For example, the tension of Abelian Higgs cosmic strings could be set by the vev of a scalar field that has an era of kination. Such a scalar can have couplings that are much larger than gravitational, realising the fast decay scenario (we do not consider the many possible, e.g., cosmological details of specific realisations of such a model).
\\
\\
In the case of reheating occurring right after the loops have decayed $t_{\rm rh}=\tau$, the amplitude can reach and exceed dark radiation bounds from BBN, as we can see by 
substituting this reheating time in Eq.~\eqref{analytic amplitude v3}, 
 \begin{equation}
     \Omega_{\rm GW, peak}^0 \simeq \Omega_{\rm loops}^i \cdot \Omega_{\rm rad}^0  \simeq 10^{-4}. \label{amplitude non-grav}
 \end{equation}
The corresponding frequency of the peak for such emission can be obtained by substituting the reheating time $t_{\rm rh}=\tau$ in Eq.~\eqref{fpeakanalytic},
\begin{align}
    f_{\rm peak}^{\rm obs} \simeq \frac{T_0}{\sqrt{\Gamma}}  \cdot \left(\frac{1}{G\mu}\right)^{3/4} \left(\frac{m_\phi}{M_P}\right)^{1/2} \simeq 10^{10} \;\text{Hz}  \cdot \left(\frac{10^{-10}}{G\mu}\right)^{3/4} \cdot \left(\frac{m_\phi}{10 \; \rm TeV}\right)^{1/2}, \label{fpeak non-grav}
\end{align}
where we have taken the upper bound for isolated loops discussed before, $\varepsilon= \sqrt{G\mu}$. If the tension $G\mu$ and the mass of the modulus $m_\phi$ depend on the volume of the compact dimensions $\mathcal{V}$ as in LVS, then $f_{\rm peak}^{\rm obs}$ is of order $T_0/\sqrt{\Gamma} \simeq  10^{10} \; \rm Hz$, independently of the value of $G\mu$ and $m_\phi$. However, given Eq.~\eqref{fpeak non-grav}, in other theories the peak can be at lower frequencies with an amplitude that is only bounded by the BBN constraint, leading to possible signals in a wider range of proposed observational searches.

\section{Conclusions}\label{sec5}

The starting point of this paper was the string loop tracker attractor solution found in \cite{SanchezGonzalez:2025uco}, in which sub-horizon cosmic string loops with time-dependent tensions constitute an order one fraction of the total energy density of the universe. As these loops decay during the subsequent moduli dominated era (before BBN or CMB formation), they produce  gravitational waves. Our aim was to compute the resulting gravitational wave signal. This is perhaps the most promising observational signature of the string loop tracker, owing to the loops' large initial energy density and the fact that their energy density redshifts like matter so is not diluted by moduli domination until it is converted to gravitational waves. The resulting spectrum is peaked at high-frequencies today (well above the MHz range), primarily because the loops must be much smaller than the Hubble parameter when they decay for the assumption that they are isolated during the tracker to hold. Notably, in this scenario the full cosmological history from soon after inflation to the present-day is consistent with known elements of string theory.
\\
\\
The amplitude of the gravitational wave signal is strongly dependent on the duration of the moduli dominated epoch. We analysed this in the context of the Large Volume Scenario (LVS) and considered three decay routes for the modulus. For standard modulus decay through gravitationally-suppressed couplings, the long era of moduli domination after the loops decay strongly suppresses the signal, as shown by Eq.~\eqref{amplitude bound grav}. 
We also considered an enhanced coupling to the Higgs sector, which is possible in the context of high-scale SUSY \cite{Cicoli:2022fzy}. This leads to a shorter moduli dominated era and a correspondingly stronger signal, as shown in Fig. \ref{fig:plotGWforLVS}. For compactification volumes (in units of $l_s^6$) $\mathcal{V}=10^4-10^{13}$ this signal can reach $\Omega_{\rm GW,peak}^0 \leq 10^{-8} - 10^{-11}$ at $f_{\rm peak}^{\rm obs} \geq 10^9 - 10^8\; \rm Hz$ respectively. Finally, if the modulus has non-gravitational couplings that allow for fast decay, the loops could survive until reheating, resulting in a signal that saturates current observational bounds on dark radiation (although it is unknown whether stringy microphysics can lead to this behaviour).
\\
\\
In future work, it would be interesting to relax the assumption that the loops are isolated. This might lead to stronger signals at lower frequencies, if larger loops were present. However, it is also plausible that the majority of the energy might be transferred to smaller loops, which would lead to the opposite effect. A further interesting direction is to connect the regime studied here (of small isolated loops) to the opposite case of dense Hagedorn phases as studied in \cite{Frey:2023khe, Frey:2024jqy, Villa:2024jbf}. The true dynamics could be determined using Nambu-Goto simulations of the string loops. Relatedly, we have modelled the distribution of loop sizes as being dominated by a single length, but our calculations could be readily extended to other distributions, if known.
\\
\\
The theories we have considered differ substantially from that usually considered in the context of gravitational waves from cosmic string networks with fixed tension. Normally, the strings are close to a scaling solution such that the network only contains a fraction of order $G\mu$ of the total energy density, however there are long strings that continually produce loops resulting in a signal that spans a wide frequency range \cite{Vilenkin:2000jqa}. Such a string network has, for example, been considered as a possible sources for the recent evidence of stochastic background signals in the nanohertz range \cite{NANOGrav:2023hvm,Figueroa:2023zhu,Ellis:2023tsl,Avgoustidis:2025svu}. There are of course many other complementary searches for cosmic string networks, for example from their possible contribution to the anisotropies in the cosmic microwave background (see e.g. \cite{Charnock:2016nzm}). As mentioned, these are not relevant for our scenario in which the string loops decay long before BBN. Interestingly, in the scenario we consider the gravitational wave signal is stronger for smaller $G\mu$, because the loops survive longer, the opposite of the trend for a network.
\\
\\
There are other ways that a stochastic gravitational wave signal at high-frequencies can be produced in the early universe. These include tachyonic preheating at the end of inflation (see e.g. \cite{Khlebnikov:1997di,Easther:2006gt,Garcia-Bellido:2007nns}), phase transitions at early times (see e.g. \cite{Athron:2023xlk} for a recent review), the decay of oscillons \cite{Zhou:2013tsa,Antusch:2016con, Amin:2018xfe}, Hagedorn Phases \cite{Frey:2024jqy}, and the Standard Model radiation bath itself if this reaches a high enough temperature after inflation \cite{Ghiglieri:2015nfa,Ghiglieri:2020mhm,Ringwald:2020ist}.
\\
\\
Meanwhile, there is a growing experimental effort targeting gravitational waves at high frequencies, see \cite{Aggarwal:2020olq, Aggarwal:2025noe} for recent reviews. Particularly relevant to the frequency range of the signals that we have considered are experiments that make use of the inverse Gertsenshtein effect, which is the conversion of gravitational waves to photons in a magnetic field \cite{Boccaletti:1970pxw}. Axion detection searches, such as IAXO, could be adapted to have sensitivity in the required frequency range; although the projected sensitivity is still from the BBN bound, quantum devices might lead to progress in the future \cite{Guo:2025cza, Berlin:2021txa}. The photons produced by the inverse Gertsenshtein effect operating in large scale, coherent magnetic fields e.g. associated with galactic clusters can also be searched for using radio telescopes (see e.g. \cite{Domcke:2020yzq} for a recent analysis). Laboratory searches using bulk acoustic wave devices are also roughly the required frequency range \cite{Goryachev:2014yra}, although again the sensitivity would need to improve dramatically to reach the BBN bound. 
\\
\\
We conclude that both the amplitude and frequency of the signal discussed here are observationally challenging. However, in the optical domain, time and technology have allowed observations that once seemed impossible. Gravitational wave astronomy sits today at the position optical astronomy was in 1620; we hope for a similarly bright future.

\acknowledgments
JC acknowledges support from the STFC consolidated grants ST/T000864/1 and \linebreak ST/X000761/1. EH acknowledges the support from the UK Research and Innovation Future Leader Fellowship MR/V024566/1. JC, NSG  and EH are also members of the COST Action COSMIC WISPers CA21106, supported by COST (European Cooperation in Science and Technology). EJC acknowledges support from the STFC Consolidated Grant ST/X000672/1. NSG acknowledges support from the Oxford-Berman Graduate Scholarship jointly funded by the Clarendon Fund and the Rudolf Peierls Centre for Theoretical Physics Studentship. For the purpose of Open Access, the authors have applied a CC BY public copyright licence to any Author Accepted Manuscript version arising from this submission.

\newpage
\bibliography{AlltheRefs}
\end{document}